# Stack Pressure Considerations for Room Temperature All-Solid-State Lithium Metal Batteries

*Jean-Marie Doux[a,#], Han Nguyen[a,#], Darren H.S. Tan[a], Abhik Banerjee[a], Xuefeng Wang[a], Erik A. Wu[a], Chiho Jo[a], Hedi Yang[a], Ying Shirley Meng[a,b]\*.*

[a] Department of NanoEngineering, University of California San Diego, La Jolla, CA 92093.

[b] Sustainable Power & Energy Center (SPEC), University of California San Diego, La Jolla, CA 92093.

\* Correspondence to: shmeng@ucsd.edu

[#] These authors contributed equally: Jean-Marie Doux, Han Nguyen.

## ABSTRACT

All-solid-state batteries are expected to enable batteries with high energy density with the use of lithium metal anodes. Although solid electrolytes are believed to be mechanically strong enough to prevent lithium dendrites from propagating, various reports today still show cell failure due to lithium dendritic growth at room temperature. While cell parameters such as current density, electrolyte porosity and interfacial properties have been investigated, mechanical properties of lithium metal and the role of applied stack pressure on the shorting behavior is still poorly understood. Here, we investigated failure mechanisms of lithium metal in all-solid-state batteries as a function of stack pressure, and conducted *in situ* characterization of the interfacial and morphological properties of the buried lithium in solid electrolytes. We found that a low stack pressure of 5 MPa allows reliable plating and stripping in a lithium symmetric cell for more than 1000 hours, and a Li | $Li_6PS_5Cl$ | $LiNi_{0.80}Co_{0.15}Al_{0.05}O_2$ full cell, plating more than 4 µm of lithium per charge, is able to cycle over 200 cycles at room temperature. These results suggest the possibility of enabling the lithium metal anode in all-solid-state batteries at reasonable stack pressures.

# INTRODUCTION

All-solid-state batteries (ASSBs) using non-flammable solid electrolytes are attracting increasing interest from their potential to enable the metallic lithium anode, which would dramatically increase energy densities compared to their liquid electrolyte counterparts. This arises from the belief that solid electrolytes serve as a suitable barrier that prevents lithium dendrite propagation.[1–3] The Monroe-Neumann criterion has postulated that a solid-state electrolyte (SSE) with a shear modulus twice that of lithium metal would be suitable to prevent such dendritic propagation.[4,5] However, internal shorting caused by lithium dendrite formation is still prevalent within SSEs that satisfy this criterion. Under ambient conditions, dendrites formed during plating are found to penetrate even near perfectly dense single crystal oxide solid electrolytes (such as $Li_7La_3Zr_2O_{12}$),[6–8] shorting the battery after a few cycles as a result.[9] Others have attempted to enable lithium metal anodes in ASSBs by increasing electrolyte density,[10] by ensuring a good wetting at the lithium-electrolyte interface,[11] or by using protective coating layers.[12] As a result, most reported literature have continued to use Li-In alloys in ASSBs.[13,14] To address this problem, recent studies have focused on the mechanical properties of lithium metal, seeking to understand the dendrite penetration mechanism within the electrolyte.[15–19] In this context, Masias *et al.* measured the Young's modulus, the shear modulus and the Poisson's ratio of lithium metal at room temperature.[15] Their findings shows that Li exhibits a yield strength of about 0.8 MPa, in accordance with former work by Tariq *et al.*,[16] over which the metal starts creeping. This has been confirmed by LePage *et al.*, which showed that at room temperature, the yield strength of Li metal is creep-dominated.[18] This yield strength needs to be correlated with the stack pressure applied on ASSBs during cycling to fully understand the mechanical behavior of a lithium metal anode. Furthermore, contrary to liquid electrolytes where optical techniques can be used to observe the morphology of the plated lithium,[20–23] observing dendrites buried inside a solid-state electrolyte requires the use of more advanced tools. Recently, Heon Kim *et al.* showed dendrites in $Li_6PS_5Cl$ electrolyte using *in situ* Auger electron spectroscopy/microscopy and scanning electron microscopy.[24] However, this requires an open cell setup, utilizing the cross-section of a cell mounted on the in-situ sample holder, which allows lithium to protrude by creeping on the edge during cycling. Seitzman *et al.* also used synchrotron X-ray microscopy to observe the formation and evolution of voids and lithium dendrites in β-$Li_3PS_4$ during plating and stripping.[25] While pressure was applied, its mechanical effects on lithium and dendrite formation was not conclusive. There is still a lack of tools capable of high-resolution morphological imaging combined with chemical species identification within *in situ* buried interfaces to identify factors causing lithium dendrite formation with solid electrolytes.

In this work, we study the influence of the applied stack pressure on the lithium metal anode in ASSBs, employing the argyrodite $Li_6PS_5Cl$ sulfide electrolyte in a closed cell setup. We first use lithium symmetric cells to determine the necessary conditions to cycle Li metal over extended durations. Then, we demonstrate stable Li metal full cell cycling over 200 cycles at room temperature. High resolution X-ray tomography and X-ray diffraction were used to observe the interfacial and morphological properties of dendrites formed during plating and stripping under higher stack pressures. Finally, we propose a mechanism for the dendrite growth in sulfide solid-state electrolytes based on the mechanical properties of Li metal.

**RESULTS & DISCUSSION**

To compare the features of typical Li metal batteries, a solid-state battery comprising $Li_6PS_5Cl$ as the solid electrolyte and $LiNbO_3$ (LNO) coated $LiNi_{0.80}Co_{0.15}Al_{0.05}O_2$ (NCA) as the cathode was used. This cell was compared against a similar one with Li-In alloy as the anode. **Figure 1** shows the voltage profiles of both cells cycled at a rate of C/10 and with a stack pressure of 25 MPa. Unlike their liquid counterparts, the SSE cannot wet new surfaces that are formed during normal battery operation; thus, high pressures are thought to be necessary to ensure consistent interfacial contact between the electrolyte and the cathode.

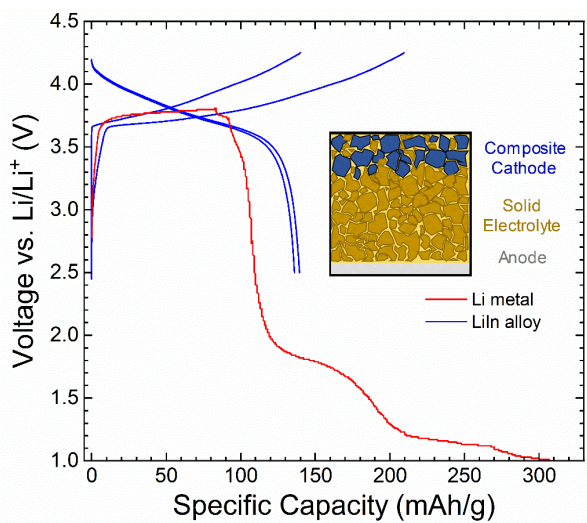

*Figure 1: First two cycles of an all-solid-state battery using a Li-In alloy anode (blue) showing typical voltage profiles, and first charge cycle using a Li metal anode (red), showing characteristic shorting behavior. Both cells were prepared in the same conditions and cycled at a stack pressure of 25 MPa.*

The cell using Li-In alloy as the anode shows an expected cycling voltage behavior with a $1^{st}$ cycle discharge capacity of approximately 140 mAh.g$^{-1}$ and a Coulombic efficiency of 66.5%. The low $1^{st}$ cycle Coulombic efficiency is attributed to initial electrolyte decomposition at the cathode.[26–28] Subsequent cycles present an average Coulombic efficiency over 99% and the cell does not exhibit any shorting behavior. In contrast, the cell using Li metal anode exhibits significant voltage drop during its $1^{st}$ charge cycle. The voltage then continues to plummet, and ultimately the cell fails to charge. This is consistent with short-circuiting behavior previously observed in the literature.[24,29] The features observed can be attributed to fundamental differences between Li metal and Li-In alloys; including: electrochemical potential, interfacial properties and mechanical properties. Although several studies have characterized the interfacial products formed between sulfide solid electrolytes and Li or Li-In alloys,[30] the differences in their mechanical properties has not been studied yet. As such, we seek to investigate this by studying the effect of stack pressure on lithium metal ASSBs.

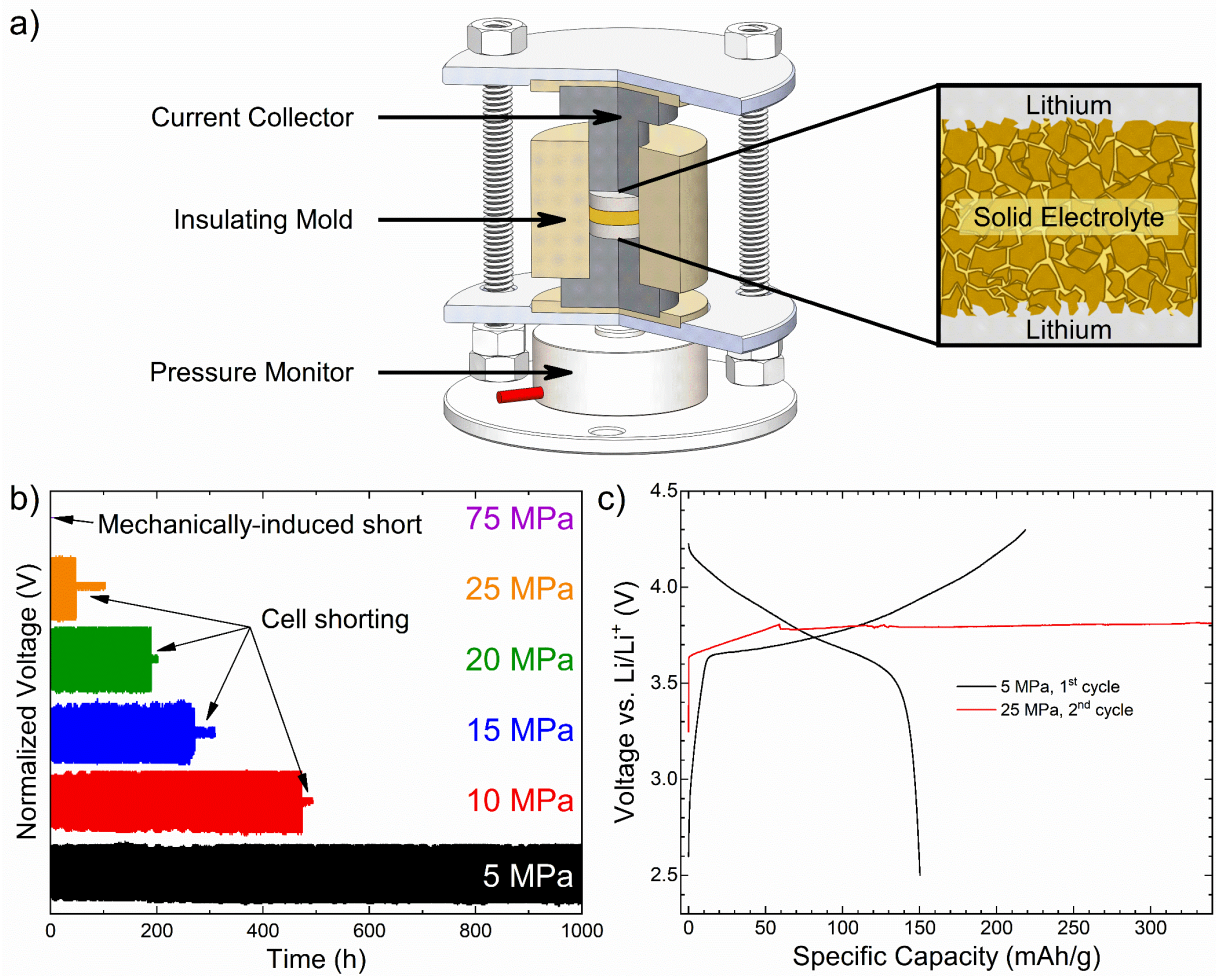

*Figure 2: a) Design of the solid-state Li symmetric cell used, allowing control and monitoring of pressure during cycling. b) Normalized voltage of Li symmetric cells as a function of time during plating and stripping at different stack pressure. At 75 MPa, the cell already mechanically shorts before cycling begins. At 5 MPa, no short was observed for over 1000 hours. c) Voltage profile of a full cell using Li metal anode: the 1$^{st}$ cycle was done at a stack pressure of 5 MPa. The stack pressure was subsequently increased on the same cell to 25 MPa before the 2$^{nd}$ cycle, during which the cell shorts.*

To investigate the effect of stack pressure on the shorting behavior of Li metal, a load cell has been added to the solid-state cell holder as shown in **Figure 2a**. The solid-state battery is then pressed between two stainless steel plates, with the bottom end in direct contact with the load cell. The stack pressure can be accurately tuned by tightening the nuts accordingly. Insulating spacers are placed between the titanium current collectors and stainless-steel plates to avoid external short-circuiting.

Next, plating and stripping of Li symmetric cells were conducted to determine stack pressure effects on dendrite formation and to determine an optimal operating pressure. **Figure 2b** shows cells plating and stripping at 75 µA.cm$^{-2}$, with continuous 1-hour plating / stripping durations until short-circuiting was observed. The cell that was initially pressed to a stack pressure of 75 MPa was observed to have shorted before the plating and stripping test began. This short circuit is therefore determined to occur mechanically and not due to any lithium plating and stripping. Since the relative density of the cold-pressed electrolyte pellets was approximately 82% (**Table S2**), it is reasonable to expect a connecting network of pores within the electrolyte. Due to the low yield strength of Li metal, creeping under such high pressure allows lithium to flow within the pores, creating an electronic percolation pathway that shorts the cell internally. When the stack pressure is lowered to 25 MPa, the symmetric cell can be cycled for approximately 48 hours before short-circuiting occurs, as indicated by a sudden overpotential drop. It is noteworthy that the cell under 25 MPa only shorts during plating and stripping. The same cell does not short when no current is applied even over prolonged durations, indicating that Li creep-induced shorting does not occur at 25 MPa. Similar tests were conducted at stack pressures of 20, 15 and 10 MPa and similar shorting behavior was observed after 190, 272 and 474 hours, respectively. The overpotentials measured in all cells were constant throughout the entire process, which indicates that stable lithium-Li$_6$PS$_5$Cl interfaces are formed. All these results show that lithium metal shorting behavior is both a mechanical and electrochemical phenomenon; a trend can be observed between stack pressure and the time needed before short-circuiting occurs.

However, at a stack pressure of 5 MPa, no short circuit was observed within 1000 hours of plating and stripping when the experiment was stopped. To confirm that this stack pressure could allow room temperature cycling of a Li metal anode, a full cell was constructed and the first cycle at a stack pressure of 5 MPa is shown on **Figure 2c**. Contrary to the Li anode cell shown previously, this battery shows a typical voltage profile without any short circuit. A specific capacity of 150 mAh.g$^{-1}$ and a first cycle Coulombic efficiency of 69% was attained, similar to the cell constructed with Li-In. In order to verify that a high stack pressure was the cause of the short circuit in **Figure 1**, the pressure was increased to 25 MPa before starting the second cycle. As seen on **Figure 2c**, a small voltage drop is observed during the charging cycle and the cell fails to charge normally. This behavior is typical of Li metal cycling and has been attributed to lithium dendrite formation during plating, generating short circuits.[24,29] These results show that the stack pressure is a crucial parameter to enable cycling of Li metal anodes in all-solid-state batteries.

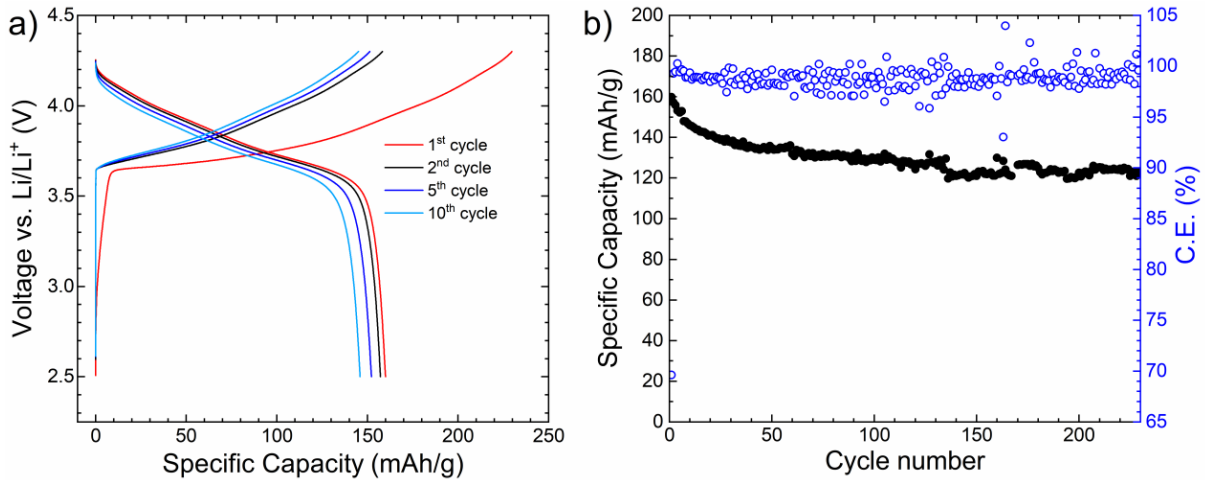

*Figure 3: a) Voltage profile of the 1st, 2nd, 5th and 10th cycle and b) cycle life of a Li metal | $Li_6PS_5Cl$ | LNO-coated NCA ASSB cycled at C/10 and at a stack pressure of 5 MPa. No shorting behavior was observed. Average Coulombic efficiency over 229 cycles is 98.86% and the cell shows a capacity retention of 80.9% over 100 cycles. The active material loading is 3.55 mg/cm².*

**Figure 3a** presents the voltage profile of the 1st, 2nd, 5th and 10th cycle of a Li metal | $Li_6PS_5Cl$ | LNO-coated NCA cell cycled at C/10 and with a stack pressure of 5 MPa, at room temperature. This cell shows stable cycling over 229 cycles (**Figure 3b**) and exhibits a capacity retention of 80.9% over 100 cycles. This demonstrates the feasibility of Li metal anodes in all-solid-state batteries. There are only a few reports of full cells cycling Li metal at room temperature in the literature, a summary is reported in **Table S1**. Unfortunately, missing experimental details in the reported literature makes the reproduction of these results difficult, and the reported cycles are limited.[24,31,32]

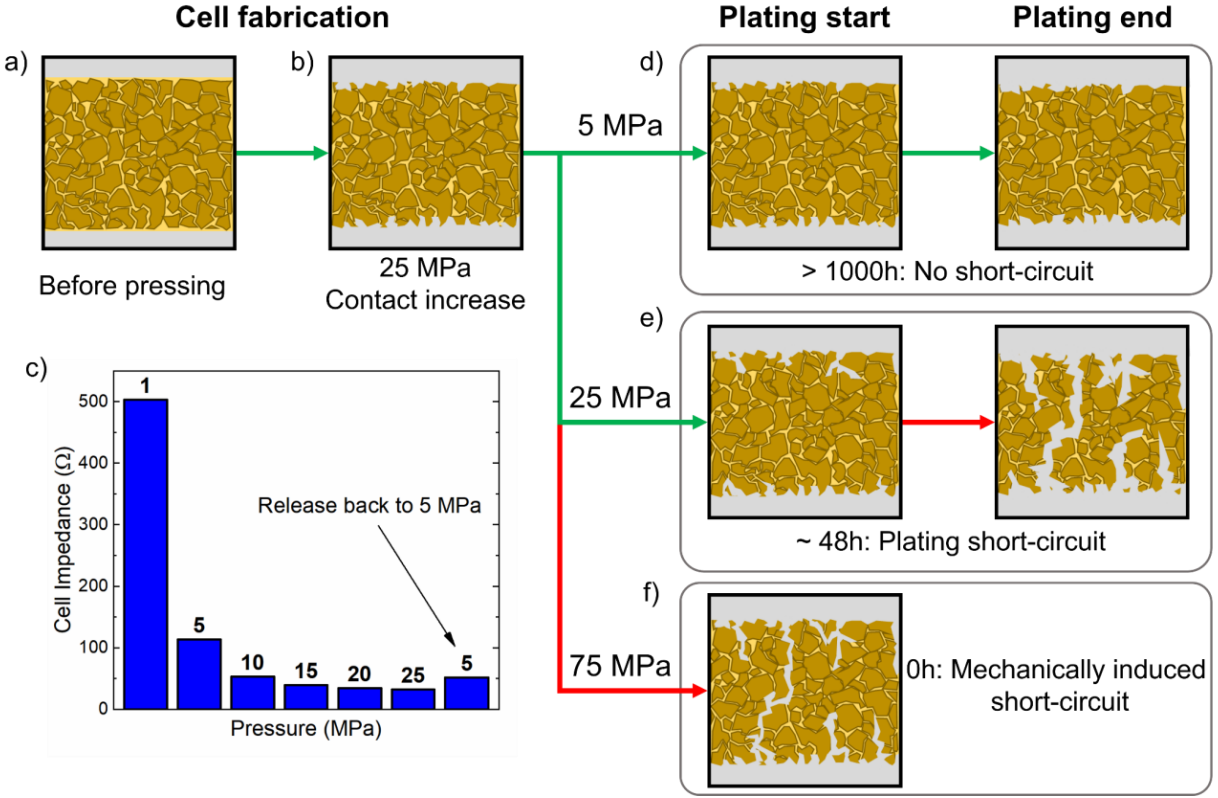

*Figure 4: Schematic of the effect of the stack pressure on the shorting behavior of Li metal solid-state batteries. a) During cell fabrication, the contact between the electrolyte and Li metal is poor before pressing the Li metal on the electrolyte pellet. b) Pressing the Li metal at 25 MPa allows for proper wetting of the electrolyte and induces a large drop in the symmetric cell impedance, as shown in c), even when the pressure is later released to 5 MPa. d) Plating and stripping at a stack pressure of 5 MPa: no creeping of Li inside the SSE pellet is observed and therefore the cell cycles for more than 1000 hours. e) At a stack pressure of 25 MPa, Li slowly creeps between the grains of the SSE and plating occurs on these dendrites, eventually shorting the cell after 48 hours. f) When the stack pressure is too high, Li creeps through the electrolyte and forms dendrites that mechanically short the cell.*

In order to understand the effect of the stack pressure on the plating and stripping of lithium in a Li symmetric cell, it is necessary to consider the creeping behavior of Li metal at each step of the cell fabrication and during cell cycling. This is detailed in **Figure 4**. First, when Li metal is added on both sides of the electrolyte pellet, interfacial contact between the two materials are poor (**Figure 4a**) and it is necessary to press the Li electrodes at 25 MPa to lower the initial cell impedance. This can be seen physically by the disappearance of voids at the interface between the Li metal and the electrolyte when using a clear polycarbonate pellet die (**Figure S1**). As shown in **Figure 4c**, the impedance of a Li metal symmetric cell depends principally of the pressure applied to improve the contact between the electrolyte and the lithium. If a pressure of only 1 MPa is used, the cell impedance exceeds 500 Ω, and this value decreases to ~110 Ω at 5 MPa, ~50 Ω at 10 MPa,

~40 Ω at 15 MPa, 35 Ω at 20 MPa, and 32 Ω at 25 MPa. Upon releasing the pressure to 5 MPa, this cell impedance only goes up to ~50 Ω, which is less than half the initial impedance at the same pressure. This can be explained by the increased wetting between lithium and the electrolyte: a relatively high pressure of 25 MPa allows lithium to creep and conforms to the relatively rough surface on the electrolyte pellet, filling the pores along the interface (**Figure 4b**). After the fabrication of the cell using a pressure of 25 MPa, three scenarios are encountered: plating and stripping at low stack pressure (5 MPa), at intermediate stack pressure (25 MPa) and at high stack pressure (75 MPa). For each case, we consider the start and end conditions of cell cycling.

For low pressure plating and stripping (**Figure 4d**), stack pressure applied is high enough to allow a good contact of lithium with the electrolyte during the cycling, but not high enough to cause lithium creep through the electrolyte and induce cell shorting. This explains why the cell shown in **Figure 2b**, when cycling under 5 MPa, was able to plate and strip for more than 1000 hours without any shorting behavior.

For a stack pressure of 25 MPa, as shown on **Figure 4e**, lithium can slowly creep inside the pores of the electrolyte to form dendrites. As the distance between the two electrodes is reduced by these small protuberances of lithium, they become the preferred sites for plating lithium due to a slightly lower overpotential experienced. Therefore, after 48 hours of plating and stripping, dendrites develop and the cell shorts, as shown previously in **Figure 2b**. Merely applying a stack pressure of 25 MPa (without any plating and stripping) did not induce any shorting, indicating that plating and stripping is necessary to form dendrites at this pressure.

Finally, when applying a stack pressure of 75 MPa (**Figure 4f**), lithium creeps through the pores of the electrolyte and the cell is shorted even before any plating and stripping starts. A porosity of 18% within the electrolyte provides connecting pathways across both electrodes, allowing lithium creep that shorts the cell. It is noted that a 75MPa stack pressure is much larger (around 100 times) than the yield strength of lithium metal.

The lowest stack pressure used (5 MPa) is still high compared to the yield strength of lithium metal (0.8 MPa) and lithium creeping could therefore be expected. Nevertheless, this yield strength value has been measured in tension, and when working in compression, the increase of the surface of contact causes the stress to gradually increase because of the friction forces.[15] A similar behavior can prevent the creeping of lithium in the ASSB at a stack pressure of 5 MPa.

For the three symmetric cell stack pressures of 5, 25, and 75 MPa, mechanical properties of the SSE pellet itself are not expected to have an influence on the shorting mechanism, as it has already been cold pressed at 370 MPa. At 75 MPa, lithium creeps through the pellet via interconnecting pores to ultimately create an electronic pathway. At 25 MPa, some lithium initially creeps into the pores but is insufficient to cause electronic short. With plating and stripping, additional force is exerted by the plated lithium deposited along the electrolyte grains, expanding the lithium filaments in the pellet until short-circuit occurs. At 5 MPa, plating of lithium only takes place on the surface of the pellet as the pressure is not high enough to allow lithium to creep into the pores.

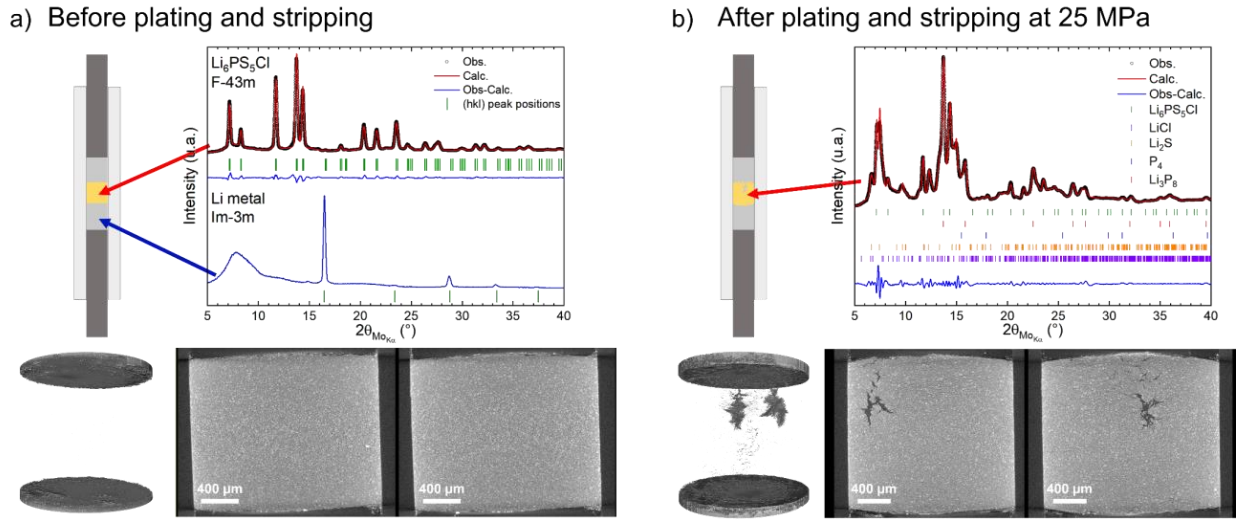

*Figure 5*: *Schematic of the cell used for X-ray tomography and X-ray diffraction, profile matching of the X-ray diffraction patterns, and X-ray tomography of a Li | $Li_6PS_5Cl$ | Li symmetric cell cycled under a stack pressure of 25 MPa a) before plating and stripping and b) after shorting. Before plating and stripping, only $Li_6PS_5Cl$ is detected in the electrolyte, and Li metal is present on both sides. The tomography pictures confirm that no lithium is present in the electrolyte. After shorting, several additional phases are detected inside the electrolyte: mainly $Li_2S$, LiCl, $P_4$ and $Li_3P_8$, all being components of the SEI formed when Li is in contact with $Li_6PS_5Cl$. Tomography pictures shows that large quantity of low-density dendrites have been formed in the electrolyte.*

Using a combination of laboratory X-ray tomography in conjunction with X-ray diffraction on the same solid-state Li symmetric cells *in situ* allowed us to obtain both morphological and chemical information of the buried dendrites. The cell was constructed with a diameter of 2 mm to allow a resolution of about 1 µm over the whole volume of the electrolyte with the X-ray tomography experiments. The use of Mo Kα radiation provided the X-ray diffraction patterns of the lithium metal and the electrolyte before and after plating and stripping. **Figure 5a** shows the X-ray tomography image of the cell before plating and stripping and XRD patterns in the lithium metal region and in the electrolyte region. The lithium metal electrodes show good contact with the electrolyte pellet; the contact interface is flat and without any voids. As expected in this pristine state, only $Li_6PS_5Cl$ is present in the electrolyte region, and the lithium metal diffraction peaks can be detected in the electrode region. After plating and stripping at 25 MPa, as shown on **Figure 5b**, the tomography images show large low-density structures within the electrolyte. These dendritic formations propagate between the grains of the electrolyte along the grain boundaries and then expand within the local sites. X-ray diffraction of the electrolyte shows the presence of numerous phases: LiCl, $Li_2S$, and reduced phosphorous species which are harder to identify because of their low concentrations. Such species have been previously identified in the literature to be the SEI formed between Li metal and $Li_6PS_5Cl$.[30] It is important to note that lithium metal dendrites are not directly detected by X-ray diffraction due to the low amounts and low scattering efficiency of lithium metal in comparison to the electrolyte and SEI products formed. Both *in situ* tomography and diffraction experiments conducted on the same cell offers direct observation of lithium dendrite growth and its corresponding interface products within the solid electrolyte. This is consistent with the proposed mechanism of cell shorting seen with electrochemical measurements discussed earlier. Although there are recent reports of void formation during stripping metallic lithium in ASSB (at 3.5 and 7 MPa) due to limited lithium – SSE contact, this issue was mitigated by improving this contact by an higher initial pressure (25 MPa) step followed by release at the working stack pressure (5 MPa). We believe that this initial high pressure step allows a homogeneous plating and stripping without formation of voids at the interface as no voids were observed by X-ray microscopy on our samples.[33]

**Conclusion**

In summary, the effect of stack pressure on the lithium metal anode in an all-solid-state battery was investigated. While stack pressure is needed to provide good initial contact between the electrolyte and the lithium by preventing the apparition of voids, a higher stack pressure can either short a cell immediately (75 MPa) or after a relatively short time of plating and stripping (25 MPa). We found that the ductility of lithium metal (due to its low stress yield) allows it to creep through the electrolyte's pores. To avoid this, a range of cycling stack pressures were studied and an optimal pressure of 5 MPa was found to allow long-term cycling of lithium metal in an all-solid-state battery. This was demonstrated in a full cell of Li | $Li_6PS_5Cl$ | NCA which cycled at room temperature for more than 200 cycles without cell failure from dendrite formation. This work paves the way toward room temperature lithium metal ASSBs and helps shed light on the importance and role of stack pressure in preventing cell failure in ASSBs.

## ASSOCIATED CONTENT

**Supporting Information**

The Supporting information contains: Experimental Details, Picture of the contact improvement after pressing the Li metal anode at 25 MPa, Literature Li metal anode ASSBs summary table, Relative density of electrolyte pellets table.

## AUTHOR INFORMATION

**Corresponding Author**

*(Y.SM.) E-mail: shirleymeng@ucsd.edu

**ORCID:**

Ying Shirley Meng: 0000-0001-8936-8845

**Notes**

## ■ ACKNOWLEDGEMENTS

This study was financially supported by the LG Chem company through Battery Innovation Contest (BIC) program. The authors would like to acknowledge UCSD Crystallography Facility. This work was performed in part at the San Diego Nanotechnology Infrastructure (SDNI) of UCSD, a member of the National Nanotechnology Coordinated Infrastructure, which is supported by the National Science Foundation (Grant ECCS-1542148). The authors would like to acknowledge the National Center for Microscopy and Imaging Research (NCMIR) technologies and instrumentation are supported by grant P41GM103412 from the National Institute of General Medical Sciences.

**Supporting Information**

# Stack Pressure Considerations for Room Temperature All-Solid-State Lithium Metal Batteries

*Jean-Marie Doux[a], Han Nguyen[a], Darren H.S. Tan[a], Abhik Banerjee[a], Xuefeng Wang[a], Erik A. Wu[a], Chiho Jo[a], Hedi Yang[a], Ying Shirley Meng[a,b]\*.*

[a] Department of NanoEngineering, University of California San Diego, La Jolla, CA 92093.

[b] Sustainable Power & Energy Center (SPEC), University of California San Diego, La Jolla, CA 92093.

\*Correspondence to: shmeng@ucsd.edu

**This SI file includes:**

Supplementary text

Fig. S1 to S6.

Tables S1, S2.

**Material & Methods**

All materials synthesis, processing, and testing was conducted in an Argon-filled glovebox (mBraun MB 200B) with $H_2O$ and $O_2$ levels below 0.1 and 0.5 ppm, respectively.

*Synthesis*

$Li_6PS_5Cl$ was purchased and used as is from NEI Corporation (USA). Li–Indium (Li-In) alloy was prepared by mixing stabilized lithium nano powder (FMC) with indium powder (Sigma Aldrich, 99.99%) with a vortex mixer until the mixture was homogeneous. The cathode $LiNi_{0.80}Co_{0.15}Al_{0.05}O_2$ (NCA, Toda Chemicals) was surface modified with an amorphous 2 wt. % $LiNbO_3$ (LNO) coating using a wet chemical method: NCA powder was added into an ethanol solution containing Li ethoxide (Sigma Aldrich) and Nb ethoxide (Sigma Aldrich) followed by stirring for an hour. The ethanol was dried in a rotary evaporator and then the powder was heated in air at 450 °C in a box furnace (Lindberg Blue M) for one hour.

*Electrochemical measurements*

All-solid-state batteries were prepared using custom-made titanium plungers and a polyether ether ketone (PEEK) die mold. 200 mg of electrolyte powder was loaded into the 13 mm PEEK die and compacted at a pressure of 370 MPa using a hydraulic press. The obtained electrolyte pellet, with a thickness of ~1 mm, has an ionic conductivity measured between 2 and 2.5 mS/cm by Electrochemical Impedance Spectroscopy (EIS).

For the Li symmetric cells, Li foil (0.5 mm thick, FMC) was cleaned with a brush to remove the surface oxidation. Li discs 12.7 mm in diameter were then cut and pressed on both side of the electrolyte pellet at a controlled pressure of 25 MPa. The cell stack pressure was then set to prescribed values of 5, 10, 15, 20, 25, 75 MPa during the plating and stripping test. Plating and stripping experiments were carried out using a 1-hour plating and stripping step, at a current of 75 $\mu A.cm^{-2}$.

For the full cells, a cathode composite was prepared with the LNO-coated NCA, $Li_6PS_5Cl$, and carbon black, in a 11:16:1 weight ratio. This mixture was mixed with an agate mortar and pestle. 12 mg of the cathode composite was then pressed on one side of the electrolyte pellet at a pressure of 370 MPa. For the anode, either a Li metal disc (12.7 mm diameter, 25 MPa) or Li-In powder (70 mg, 120 MPa) was pressed on the other side of the electrolyte pellet. Cells were cycled at a rate of C/10 between 2.5 and 4.3V vs. $Li/Li^+$.

All cells were cycled inside an Ar-filled glovebox using Landhe battery cyclers.

*Pressure monitoring and control*

Control of the stack pressure applied to the solid-state batteries during cycling was conducted with a special cell holder (detailed in Figure 1). A 3-ton load cell is mounted at the bottom of the holder, in the axis of the battery, and allows measurement of the pressure between 0 and 220 MPa. The load cell was calibrated using a 100 kN Instron 5982 Universal Testing System mechanical testing frame by applying a known load at 100% of the capacity of the load cell. Linearity of the calibration was then verified over the whole range of the load cell.

*X-ray Tomography*

For the X-ray tomography measurements, a specially designed cell was used to allow higher resolution (i.e. small voxel size). A 3.2 mm diameter polycarbonate rod was bored at 2 mm internal diameter and 2 mm steel plungers were used to fabricate and cycle the cell. Around 5 mg of electrolyte ($Li_6PS_5Cl$) was pelletized at a pressure of 370 MPa, corresponding to a pellet thickness of around 2 mm. Lithium metal strips were then cut from a cleaned Li foil and pressed on both sides of the pellet at 25 MPa. Enough Li metal was used to ensure that there was more than 5 mm of Li on each side of the cell in order to prevent artifacts caused by the high absorption of the steel plungers on the X-ray microscopy pictures.

X-ray computer-assisted tomography was done with a Versa 510 (Zeiss/Xradia) X-ray microscope, with a source voltage of 80 kV and a power of 6.5 W, using the LE2 filter. A 4X objective was used with an exposure of 0.5 s and 1601 projections, giving a resolution of about 1.18 µm. Analysis of the reconstruction data was performed using Amira 2019.1 (ThermoFisher Scientific) and Fiji software.[1,2]

*X-ray Diffraction*

X-ray diffraction experiments were performed on the solid-state Li symmetric cells used for the X-ray tomography experiments. As Cu Kα radiation are not energetic enough to work in transmission mode with this cell design, Mo Kα radiation ($\lambda = 0.709320$ Å) was used. A Bruker Apex II Ultra diffractometer was used in transmission mode equipped with a 2D CCD detector.

Refinement of the diffraction patterns were carried out in profile matching mode using the FullProf software suite.[3]

**SI figures**

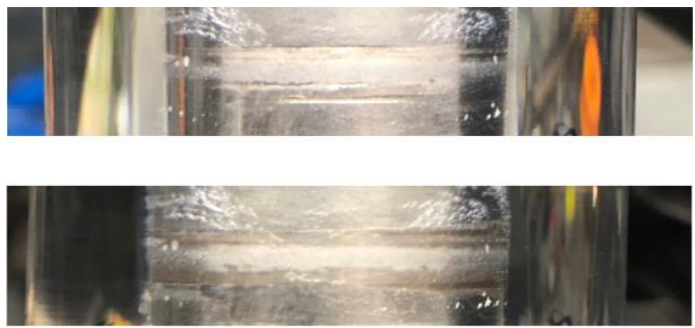

*Figure S1: Pictures of a Li metal | $Li_6PS_5Cl$ | Li metal symmetric cell before (top) and after (bottom) pressing at a pressure of 25 MPa. The contact improvement between the Li metal and the electrolyte pellet can be seen by the disappearance of voids at the interface.*

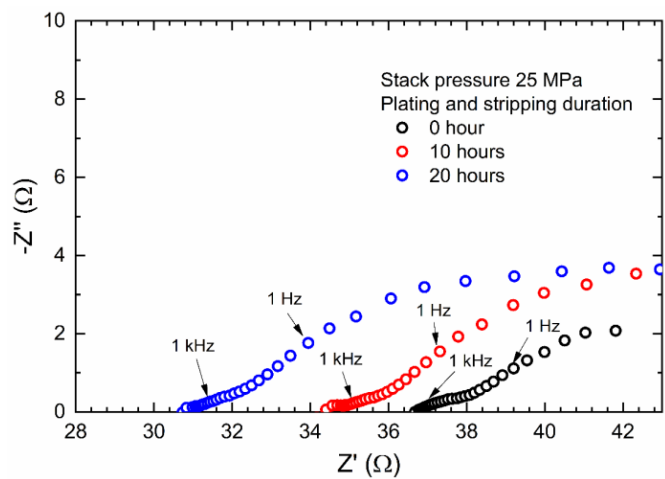

*Figure S2: Nyquist diagram of a Li symmetric cell before and after plating and stripping for 10 and 20 hours, at a stack pressure of 25 MPa.*

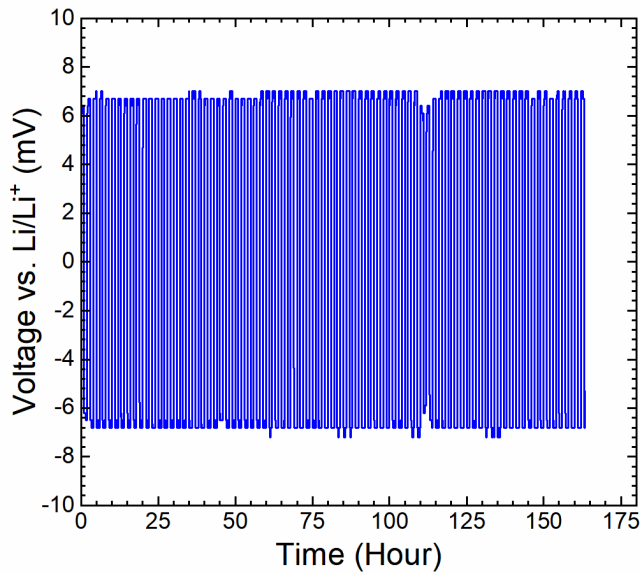

*Figure S3: Potential of a Li symmetric cell during plating and stripping at a stack pressure of 2 MPa.*

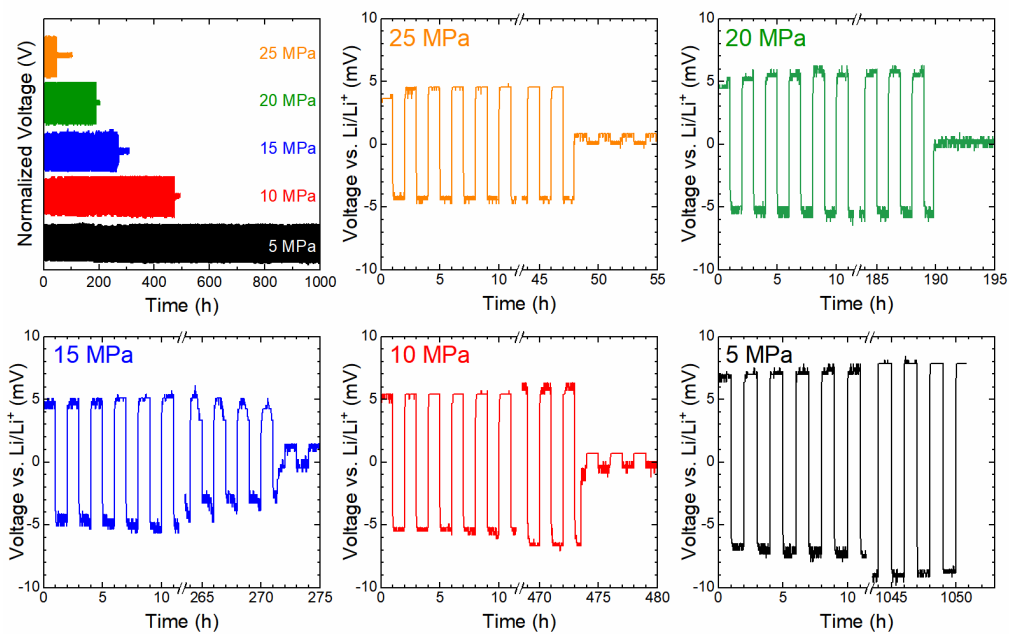

*Figure S4: Voltage profiles of the Li symmetric cells during plating and stripping at different stack pressures.*

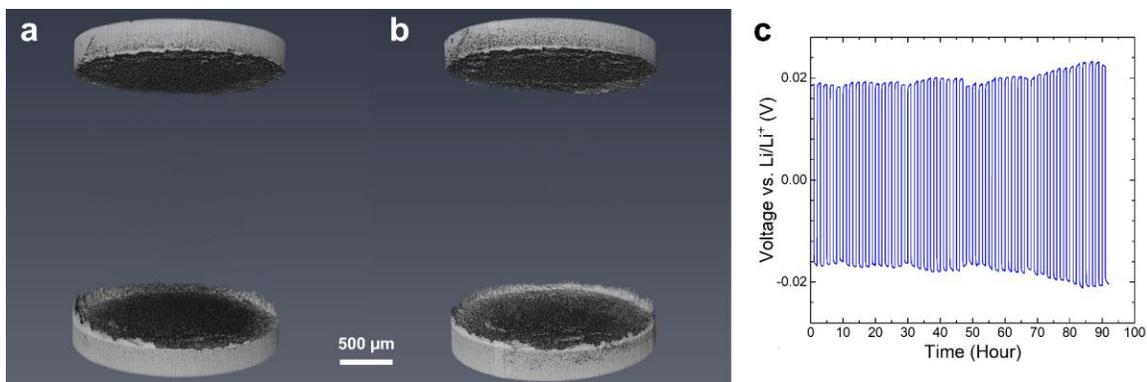

*Figure S5: X-ray tomography of a Li symmetric cell before (a) and after (b) 92 hours of plating and stripping at a stack pressure of 5 MPa. c) Voltage profile during the plating and stripping at 5 MPa.*

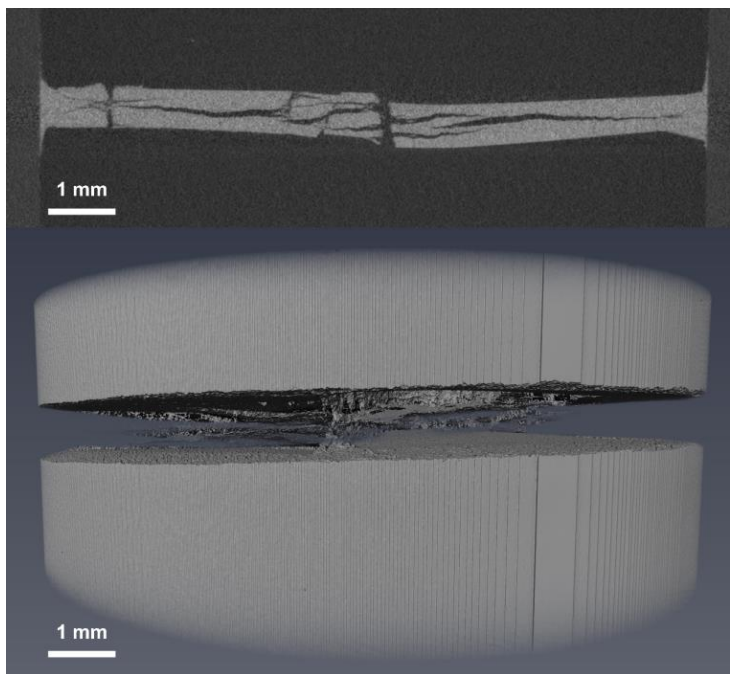

*Figure S6: X-ray tomography of a Li symmetric cell mechanically shorted after pressing Li metal at 75 MPa.*

*Table S1: Summary of the reported Li metal cycling in ASSBs in the literature.*

| Cathode | Electrolyte | Number of Cycles | Charge/Discharge Rate | Current Density | Capacity retention | Reference |
|---|---|---|---|---|---|---|
| $LiNi_{0.8}Mn_{0.05}Co_{0.15}O_2$ | $Li_6PS_5Cl$ | 100 | 0.22C / 0.35C | / | 74% - 100 cycles | [5] |
| $LiCoO_2$ | $Li_2S-Li_2O-P_2S_5$ | 25 | / | 50 µA/cm² | ~50% - 25 cycles | [6] |
| $Li_2S$ | $Li_3PS_4$ | 100 | 0.2C / 0.2C | / | ~79% - 100 cycles | [7] |

*Table S2: Relative density calculation based on physical measurements of four $Li_6PS_5Cl$ electrolyte pellets, using a theoretical density value of 1.860 g.cm$^{-3}$.[4]*

| Mass (mg) | Diameter (mm) | Thickness (mm) | Exp. Density (g.cm$^{-3}$) | Relative Density (%) |
|---|---|---|---|---|
| 200.0 | 13 | 1.01 | 1.492 | 80.2 |
| 201.2 | 13 | 0.96 | 1.579 | 84.9 |
| 200.3 | 13 | 1.01 | 1.494 | 80.3 |
| 200.3 | 13 | 0.99 | 1.524 | 83.0 |
| | **Average** | | **1.522** | **82.1** |